\documentclass[twocolumn,showpacs,preprintnumbers,amsmath,amssymb,superscriptaddress]{revtex4}

\usepackage{graphicx}
\usepackage{dcolumn}
\usepackage{bm}
\usepackage{amsmath}

\newcommand{\be}{\begin{equation}}
\newcommand{\ee}{\end{equation}}
\newcommand{\beq}{\begin{eqnarray}}
\newcommand{\eeq}{\end{eqnarray}}

\def\epee{\mathrel{{\epsilon_{ee}}}}
\def\epem{\mathrel{{\epsilon_{e\mu}}}}
\def\epmm{\mathrel{{\epsilon_{\mu\mu}}}}
\def\epet{\mathrel{{\epsilon_{e\tau}}}}
\def\epmt{\mathrel{{\epsilon_{\mu\tau}}}}
\def\eptt{\mathrel{{\epsilon_{\tau\tau}}}}

\def\nue{\mathrel{{\nu_e}}}
\def\numu{\mathrel{{\nu_\mu}}}
\def\nutau{\mathrel{{\nu_\tau}}}

\def\t13{\mathrel{{\theta_{13}}}}
\def\y12{\mathrel{{\tan^2 \theta_{12}}}}

\begin{document}

\preprint{LA-UR-04-5646, YITP-SB-04-43}

\title{Atmospheric neutrinos as probes of neutrino-matter interactions}

\author{Alexander Friedland}
 \email{friedland@lanl.gov}
\affiliation{%
Theoretical Division, T-8, MS B285, Los Alamos National Laboratory, Los Alamos, NM 87545
}%

\author{Cecilia Lunardini}
 \email{lunardi@ias.edu}
\affiliation{
School of Natural Sciences, Institute for Advanced Study, Einstein Drive, Princeton, NJ 08540
}%
\author{Michele Maltoni}
 \email{maltoni@insti.physics.sunysb.edu}
\affiliation{
C.~N.~Yang Institute for Theoretical Physics,
SUNY at Stony Brook, Stony Brook, NY 11794-3840
}%

\date{August 24, 2004}

\begin{abstract}
  Neutrino oscillation experiments provide a unique tool for probing
  neutrino-matter interactions, especially those involving the tau
  neutrino. We describe the sensitivity of the present atmospheric
  neutrino data to these interactions in the framework of a
  three-flavor analysis. Compared to the two-flavor $\nu_\mu-\nu_\tau$
  analyses, we find qualitatively new features, in particular, that
  large non-standard interactions, comparable in strength to those in
  the Standard Model, can be consistent with the data. The existence
  of such interactions could imply a smaller value of the neutrino
  mixing angle and larger value of the mass-squared splitting than in
  the case of standard interactions only.  This and other effects of
  non-standard interactions may be tested in the next several years by
  MINOS, KamLAND and solar neutrino experiments.
\end{abstract}

\maketitle

\section{Introduction}
Historically, the studies of the neutrino properties have lead to
several important breakthroughs in particle physics. The discovery
of the neutral current processes played a key role in establishing
the Standard Model (SM), while the recent confirmation of neutrino
oscillations provided the first real sign of non-standard physics.
To this day, however, neutrino interactions remain one of the
least tested aspects of the SM. The importance of their direct
experimental determination, whether it leads to further
confirmation of the SM or to a discovery of new
physics, can hardly be overstated.

In this paper, we explore the possibility of measuring
neutrino-matter interactions using the atmospheric neutrino data.
A key feature of these data is their sensitivity to the
interactions of the tau neutrino.  Indeed, atmospheric muon
neutrinos are believed to oscillate into tau neutrinos; modified
interactions of the latter may make the oscillation pattern
incompatible with the data.  In contrast, accelerator-based
experiments have restricted mainly the interactions of the muon
neutrino.

The sensitivity of the atmospheric neutrino data to non-standard
interactions (NSI) was studied in
\cite{Maltoni2001,Guzzo2001,ConchaMichele2004}. It was found that, in
the context of a two-flavor $\nu_\mu\leftrightarrow\nu_\tau$ analysis,
these data are quite sensitive to NSI. The sensitivity to possible
$\nu_e\leftrightarrow\nu_\tau$ conversion, however, was not
systematically explored beyond some specific examples
\cite{Guzzo2001}.  Solar neutrino data constrain this channel
\cite{oursolar}, but at the same time leave room for non-trivial
effects, both on the solar and atmospheric neutrinos. In the solar
case, the $\nu_e\leftrightarrow\nu_\tau$ interaction gives rise to a
new disconnected solution LMA-0 \cite{oursolar},
\cite{brazilians,valencians}, thus introducing an uncertainty in the
determination of the solar oscillation parameters. Whether a combined
analysis of the solar and atmospheric data eliminates this uncertainty
is an important outstanding question.

In the following, we perform the first study of the sensitivity of the
atmospheric data to NSI in the three-flavor framework. Our goals are:
first, to describe the physics behind the sensitivity of the data to
NSI; second, to present the bounds on the interaction parameters,
based on a detailed fit to the data; and finally, to discuss how the
possible presence of modified interactions can affect the measurement
of the oscillation parameters.  We intend to publish the results of a
complete scan of the parameter space elsewhere \cite{inprep}.

\section{Setup}
The atmospheric neutrino flux is comprised of both neutrinos and
antineutrinos, with the energy spectrum spanning roughly five
orders of magnitude, from $10^{-1}$ to $10^4$ GeV
\cite{EngelPLB2000}. Initially created as the electron and muon
flavor states, these neutrinos (depending on the distance
traveled) undergo flavor oscillations. The oscillations are
governed by the Hamiltonian $H$, which has a kinetic (vacuum) part
and an interaction part. The former is given, at the leading
order, by
\begin{equation}
  \label{eq:vacHprime}
    H_{\rm vac}\simeq\Delta \left(\begin{array}{ccc}
   - 1 & 0 & 0 \\
   0 & -\cos 2\theta & \sin 2\theta \\
  0  & \sin 2\theta & \cos 2\theta \\
\end{array}\right),
\end{equation}
where we omit the smaller, ``solar'' mass splitting, $\Delta
m_\odot^2$ \cite{solareffects}, and the remaining two mixing angles,
$\theta_{12}$ and $\theta_{13}$.  In Eq.~(\ref{eq:vacHprime}),
$\Delta\equiv \Delta m^2/(4 E_\nu)$, $E_\nu$ is the neutrino energy
and $\Delta m^2$ is the largest mass square difference in the neutrino
spectrum; the flavor basis, $ \nue,\numu,\nutau $, is assumed.  The
interaction part (up to an irrelevant constant) will be parameterized as
\begin{equation}
H_{\rm mat}=\sqrt{2}G_F n_e\begin{pmatrix}
1+\epee & \epsilon_{e\mu}^\ast & \epsilon_{e\tau}^\ast \\
\epem & \epmm& \epsilon_{\mu\tau}^\ast \\
\epet & \epmt & \eptt \\
\end{pmatrix}.
\label{eq:ham}
\end{equation}
Here $G_F$ is the Fermi constant, $n_e$ is the number density of
electrons in the medium.

The physical origin of the epsilon contributions in $H_{\rm mat}$ can be
the exchange of a new heavy vector or scalar \footnote{
Scalar
  interactions of the form $(\bar{\nu}f_R)(\bar{f}_R\nu)$ reduce to
  Eq.~(\ref{eq:lagNSI}) upon the application of the Fierz
  transformation, see e.g. S.~Bergmann, Y.~Grossman and E.~Nardi,
Phys.\ Rev.\ D {\bf 60}, 093008 (1999)
[arXiv:hep-ph/9903517].
}
particle. We parameterize the resulting NSI with
the effective low-energy four-fermion Lagrangian
\begin{eqnarray}
L^{NSI} &=& - 2\sqrt{2}G_F (\bar{\nu}_\alpha\gamma_\rho\nu_\beta)
(\epsilon_{\alpha\beta}^{f\tilde{f} L}\bar{f}_L \gamma^\rho
\tilde{f}_L + \epsilon_{\alpha\beta}^{f\tilde{f}
R}\bar{f}_R\gamma^\rho \tilde{f}_{R})\nonumber\\
&+& h.c. \label{eq:lagNSI}
\end{eqnarray}
Here $\epsilon_{\alpha\beta}^{f\tilde{f} L}$
($\epsilon_{\alpha\beta}^{f\tilde{f} R}$) denotes the strength of the
NSI between the neutrinos $\nu$ of flavors $\alpha$ and $\beta$ and
the left-handed (right-handed) components of the fermions $f$ and
$\tilde{f}$. The epsilons in Eq.~(\ref{eq:ham}) are \emph{the sum} of
the contributions from electrons ($\epsilon^{e}$), up quarks
($\epsilon^{u}$), and down quarks ($\epsilon^{d}$) in matter:
$\epsilon_{\alpha\beta}\equiv
\sum_{f=u,d,e}\epsilon_{\alpha\beta}^{f}n_f/n_e$. In turn,
$\epsilon_{\alpha\beta}^{f}\equiv\epsilon_{\alpha\beta}^{fL}+\epsilon_{\alpha\beta}^{fR}$
and $\epsilon_{\alpha\beta}^{fP}\equiv\epsilon_{\alpha\beta}^{ffP}$.
Notice that the matter effects are sensitive only to the interactions
that preserve the flavor of the background fermion $f$ (required by
coherence \cite{nunu}) and, furthermore, only to the vector part of
that interaction.

Neutrino scattering tests, like those of NuTeV \cite{nutev} and CHARM
\cite{charm}, mainly constrain the NSI couplings of the muon neutrino,
e.g., $|\epem|\lesssim 10^{-3}$, $|\epmm|\lesssim 10^{-3}-10^{-2}$. The
limits they place on $\epee$, $\epet$, and $\eptt$ are rather loose,
e. g., $|\epsilon_{\tau\tau}^{uu R}|<3$, $-0.4<\epsilon_{ee}^{uu
  R}<0.7$, $|\epsilon_{\tau e}^{uu}|<0.5$, $|\epsilon_{\tau
  e}^{dd}|<0.5 $ \cite{Davidson:2003ha}. Stronger constraints exist on
the corresponding interactions involving the charged leptons. Those,
however, cannot, in general, be extended to the neutrinos, for example
when the underlying operators contain the Higgs fields
\cite{BerezhianiRossi}, and hence will not be considered here.

Given the above bounds we will set $\epem$ and $\epmm$ to zero in our
analysis. Furthermore, for simplicity we will also set $\epmt$ to
zero. The earlier analyses of the atmospheric neutrino data
\cite{Maltoni2001} have indicated that this parameter is quite
constrained ($\epmt<10^{-2}-10^{-1}$). Corrections due to non-vanishing
$\epmt$ will be described in \cite{inprep}. Here, we have a
three-dimensional NSI parameter space, spanned by $\epee$, $\epet$,
and $\eptt$.

\section{Conversion effects and sensitivity to the NSI}
The physics of the sensitivity of the atmospheric neutrino data to
$\epee$, $\epet$, and $\eptt$ can be understood as follows. The data
are known to be very well fit by large-amplitude oscillations between
the $\nu_\mu$ and $\nu_\tau$ states. This holds both at high energy
($E_\nu\gtrsim 10$ GeV), where only the muon neutrino flux is
measured, and at lower energies, where both the muon and electron
neutrino data are available. These oscillations are driven by the
off-diagonal $\nu_\mu-\nu_\tau$ mixing in Eq.~(\ref{eq:vacHprime}) and
the introduction of sufficiently large NSI for the tau neutrino will,
in general, suppress that mixing. Since the vacuum Hamiltonian scales
as $E_\nu^{-1}$, this suppression should be especially strong at high
energy, in the through-going muon sample.

As a simple illustration, consider the case when only $\eptt$ is
nonzero.  Clearly, $\eptt$ introduces a diagonal splitting between the
$\nu_\mu$ and $\nu_\tau$ states, thereby decreasing the effective
mixing angle in matter.  The corresponding bound can be estimated by
comparing the matter term $\sqrt{2}\eptt G_F n_e$ to the vacuum
oscillation term $\Delta m^2/(2 E_\nu)$. For neutrinos going through
the center of the Earth, the highest energy at which an oscillation
minimum occurs in the standard case is around $E_0 \sim 20-30$ GeV. If
the matter term is sufficiently large, $\sqrt{2}\eptt G_F n_e\gtrsim
\Delta m^2/(2 E_0)$, the mixing in matter and hence the oscillation
amplitude are expected to be suppressed. Substituting numerical
values, we find a bound $\eptt\lesssim 0.2$.

Next, we generalize this argument to the case of non-vanishing
$\epee$, $\epet$. The matter part of the Hamiltonian $H_{\rm mat}$ can
be diagonalized by rotating in the $\nu_e-\nu_\tau$ subspace. In the
new basis $(\nu_{e'},\nu_\mu,\nu_{\tau'})$, $H_{\rm mat}$ has the form
$diag(\lambda_{e'},0,\lambda_{\tau'})$, with
$\lambda_{e',\tau'}=\sqrt{2}G_F n_e
(1+\epee+\eptt\pm\sqrt{(1+\epee-\eptt)^2+4 |\epet|^2})/2$. It
straightforwardly follows that if $|\lambda_{e',\tau'}|\ll \Delta
m^2/(2 E_0)$, the oscillations of the muon neutrinos proceed
unimpeded, while in opposite case, $|\lambda_{e',\tau'}|\gtrsim \Delta
m^2/(2 E_0)$, they are suppressed.

It is very important to consider the intermediate regime, when the
spectrum has the hierarchy (a) $|\lambda_{\tau'}|<\Delta m^2/(2
E_0)\ll|\lambda_{e'}|$ or (b) $|\lambda_{e'}|<\Delta m^2/(2
E_0)\ll|\lambda_{\tau'}|$. In both cases, the oscillations between
$\nu_\mu$ and the corresponding light eigenstate are allowed to
proceed while those between $\nu_\mu$ and the heavy eigenstates
are suppressed.  Remarkably, the resulting oscillation pattern is
\emph{indistinguishable} from the standard case at high energy,
where only muon neutrinos are detected.

From now on we specialize to hierarchy (a), which is smoothly
connected to the origin $\epee=\epet=\eptt=0$ ((b) is realized
only if $\epee+\eptt$ is a large negative number). When it is
satisfied, muon neutrinos oscillate into the state
\begin{equation}
  \label{eq:tauprime}
  \nu_{\tau'}=-s_\beta e^{2 i \psi} \nue + c_\beta  \nutau~,
\end{equation}
where $c_\beta = \cos \beta$,  $s_\beta = \sin \beta$,
$ 2 \psi = Arg(\epet)$, $\tan 2\beta = 2 | \epet|/( 1+\epee -\eptt)$.

The condition $|\lambda_{\tau'}|\lesssim \Delta m^2/(2 E_0)$ implies
\begin{equation}
  \label{eq:width}
  |1+\epee+\eptt-\sqrt{(1+\epee-\eptt)^2+4 |\epet|^2}|\lesssim 0.4.
\end{equation}
This equation gives our analytical prediction for the bound on
$\epee, \epet, \eptt$. When $\epee=\epet=0$, it reduces to the
bound $\eptt\lesssim 0.2$ given above.

The region Eq.~(\ref{eq:width}) describes extends to large values of
$\epet$, $\eptt$. To see this, note that in the limit
$\lambda_{\tau'}=0$, or
\begin{equation}
  \label{eq:parabola}
  \eptt = |\epet|^2/(1+\epee)~,
\end{equation}
the $\numu \leftrightarrow \nu_{\tau'}$ oscillations, though dependent on
the matter angle $\beta$, are independent of the absolute size of the
NSI. As already mentioned, these oscillations have the same dependence
on the neutrino energy and on the distance $L$ as vacuum oscillations
and therefore mimic their effect in the distortion of the neutrino
energy spectrum and of the zenith angle distribution.  More
specifically, we get the oscillation probability:
 \be
  P(\numu\rightarrow \nu_{\tau'})=
  \sin^2 2\theta_m \sin^2 [ \Delta
    m^2_m L/(4 E_\nu) ]~,
 \label{prob}
 \ee
where the effective mixing and mass square splitting are derived
to be
 \beq
  &&\Delta m^2_{m} = \Delta m^2 \left[
  (c_{2\theta} (1+ c^2_\beta) - s^2_\beta)^2/4 + (s_{2\theta}
  c_{\beta})^2\right]^{1/2} ~, \nonumber \\
  &&\tan 2\theta_m = 2
  s_{2\theta} c_\beta/(c_{2\theta} (1+ c^2_\beta) - s^2_\beta )~.
 \label{effmix}
 \eeq
If NSI are present, but not included in the data analysis, a fit
of the highest energy atmospheric data, i.e. the through-going
muon ones, would give $\Delta m^2_{m} $ and $\theta_m $ instead of
the corresponding vacuum quantities. If we fix a set of NSI and --
to reproduce the no-NSI case -- require that $\theta_m \simeq
\pi/4$ and $\Delta m^2_m\simeq 2.5 \cdot 10^{-3}~{\rm eV^2} $,
from Eqs. (\ref{effmix}) we get that the vacuum mixing would {\it
not} be maximal; in particular we have $\cos 2\theta \simeq
s^2_\beta /(1+ c^2_\beta)$ and $\Delta m^2 \simeq \Delta
m^2_m(1+\cos^{-2}\beta)/2$.

In the intermediate energy range, $E \sim 1 - 10$ GeV, when matter
and vacuum terms are comparable, the reduction to a two-neutrino
system is not possible, and the problem does not allow a simple
analytical treatment. The neutrino conversion probability in this
energy range depends on the sign of the neutrino mass hierarchy
(normal, $\Delta m^2>0$, or inverted, $\Delta m^2<0$).  At the
sub-GeV energies, we expect vacuum-domination, and therefore small
deviations with respect to vacuum oscillations \footnote{The
matter effects may
  give beat-like structure to the survival probability as a function
  of energy $E_\nu$, which tends to become unobservable upon
  integration over $E_\nu$.}.

Finally, we observe that for $\theta_{13}= 0$, $\Delta m_\odot= 0$,
as has been assumed here, there is no sensitivity to $\psi$, the
phase of $\epet$ \footnote{Clearly, by redefining
the phases of $\nu_\mu$ and $\nu_\tau$ one changes the phases
of the $e\mu$ and $e\tau$ entries of the Hamiltonian.}. This is
unlike the case of the solar neutrinos, where $\psi$ plays a
crucial role \cite{oursolar}.  Corrections due to $\theta_{13}$ and
$\Delta m_\odot\ne 0$
break the phase degeneracy and will be presented elsewhere
\cite{inprep}.

\section{Numerical results}
We performed a quantitative analysis of the atmospheric neutrino data
with five parameters: two ``vacuum" ones, $(\Delta m^2,\theta)$, and
three NSI quantities $(\epee,\epet,\eptt)$. The goodness-of-fit for a
given point is determined by performing a fit to the data. We use the
complete 1489-day charged current Super-Kamiokande phase I data
set~\cite{Hayato:2003}, including the $e$-like and $\mu$-like data
samples of sub- and multi-GeV contained events (each grouped into 10
bins in zenith angle) as well as the stopping (5 angular bins) and
through-going (10 angular bins) upgoing muon data events. This amounts
to a total of 55 data points. For the calculation of the expected
rates we use the new three-dimensional atmospheric neutrino fluxes
given in Ref.~\cite{Honda:2004yz}. The statistical analysis of the
data follows the appendix of Ref.~\cite{ConchaMichele2004}.

The results of the K2K experiment have been included.
Their addition has a minimal impact on our
results, providing some constraint at high $\Delta m^2$.  The details
of the K2K analysis can be found in Sec.~2.2 of
Ref.~\cite{Maltoni:2004ei}.

\begin{figure}[htbp]
  \centering
  \includegraphics[width=0.47\textwidth]{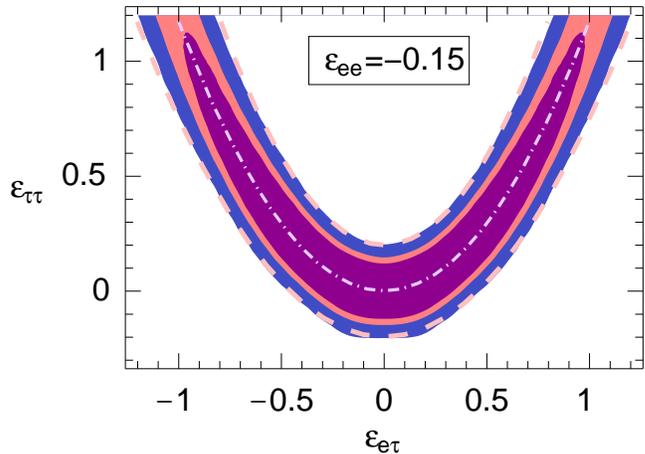}
  \caption{A 2-D section $(\epee=-0.15)$ of the allowed region
  of the NSI parameters (shaded). We assumed $\Delta m^2_\odot=0$ and
  $\theta_{13}=0$, and marginalized over $\theta$ and $\Delta m^2 $.
  The dashed contours indicate our
  analytical predictions. See text for details.}
  \label{fig:scan}
\end{figure}

Upon scanning the parameter space and marginalizing over $\Delta m^2$
and $\theta$ we obtain the three-dimensional allowed region in the space
$(\epee,\epet,\eptt)$.
As an illustration, in Fig.~\ref{fig:scan} we show a section of this
region by the plane $\epee=-0.15$ (the choice motivated by the solar
analysis in \cite{oursolar}). The $\chi^2$ minimum occurs at
$\epet=0.07,\eptt=0.01$; the value at the minimum, $\chi^2_{\rm
  min}=48.50$, is virtually the same as at the origin (no NSI),
$\chi_{\rm orig}^2=48.57$. The shaded regions correspond, from the innermost contour, to
$\chi^2-\chi^2_{\rm min}\leq$ 7.81, 11.35, and 18.80.
They represent the 95\%, 99\%, and
$3.6\sigma$ confidence levels (C.L.)  for three degrees of freedom
(d.o.f.). The last contour also corresponds to the 95\% C.L. for
50 d.o.f..  For the purpose of hypothesis
testing this means that a theory which gives  NSI outside of this
region should be rejected.

The dashed-dotted parabola illustrates the condition of zero
eigenvalue, Eq.~(\ref{eq:parabola}); the two outer curves give the
predicted bound according to Eq.~(\ref{eq:width}). For both, the
agreement between the theory and numerical results is quite
convincing. Moreover, we have verified that the agreement remains
very good for $\epee$ in the range $-0.7 < \epee < 0.3$
\cite{inprep}. For the case when only $\epet$ is non-zero we find
the bounds $|\epet| < 0.38$ at 99\% C.L. and $|\epet| < 0.5$ at
$3.6\sigma$.

The extent of the allowed region \emph{along the parabola} is beyond
the scope of our analytical treatment.  Indeed, since at high energy
the leading NSI effect is canceled by construction, the fit quality is
determined by subdominant NSI effects in all energy samples.
Remarkably, these effects are rather small, especially for the
inverted mass hierarchy, where the region
$\chi^2-\chi^2_{\rm min}\leq 7.81$ extends up to $\epet=1$, as the
Figure shows.  The extent is somewhat smaller (to $\epet=\pm 0.6$) for
the normal hierarchy \cite{inprep}.

As mentioned, the symmetry of Fig.~\ref{fig:scan} with respect
to $\epet\leftrightarrow-\epet$ follows from setting
$\theta_{13}$ and $\Delta m_\odot^2$ to zero. We have checked that for
$\theta_{13}\ne 0$ the region does indeed develop a small asymmetry
\cite{inprep}.

\begin{figure}[tbp]
  \centering
  \includegraphics[width=0.47\textwidth]{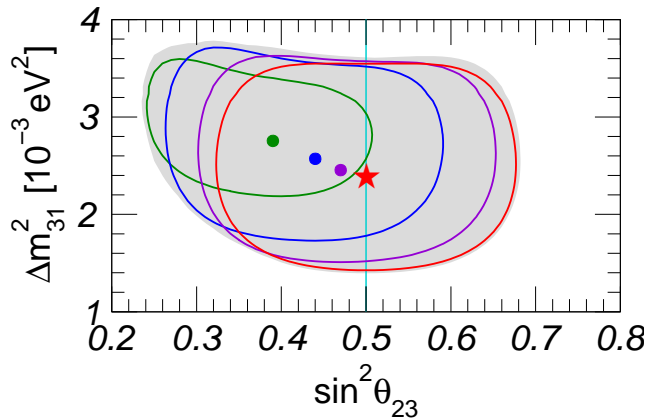}
  \caption{The effect of the NSI on the allowed region and best-fit values of the
  oscillation parameters; see text for details.}
  \label{fig:osc_marginalized}
\end{figure}

Finally, we notice that as the NSI point is moved along the parabola
(\ref{eq:parabola}) away from the origin, the best-fit oscillation
parameters change. This behavior is illustrated in
Fig.~\ref{fig:osc_marginalized}.  The four contours plotted are the
sections of the five-dimensional allowed region in the space $(\Delta
m^2,\theta,\epee,\epet,\eptt)$ for the values: (i)
$\epee=-0.15,\epet=0,\eptt=0$; (ii)
$\epee=-0.15,\epet=0.30,\eptt=0.106$; (iii)
$\epee=-0.15,\epet=0.60,\eptt=0.424$; (iv)
$\epee=-0.15,\epet=0.90,\eptt=0.953$. The contours are drawn for
$\Delta\chi^2=11.07$ with respect to the \emph{global} minimum, marked
by a star (this corresponds to 95\% CL for 5 d.o.f.). The dots
represent the best fit points obtained in each of the four cases
above, when the NSI are assumed known.  The figure confirms the trend
predicted in Eq.~(\ref{effmix}): as the NSI increase the best-fit
mixing angle becomes smaller, $\theta<\pi/4$, while the best-fit
$\Delta m^2$ increases.

The shaded region in the background represents the result of
marginalizing with respect to $\epet$ and $\eptt$ (for
$\epee=-0.15$). It was drawn for $\Delta\chi^2=11.83$,
corresponding to 99.73\% ($3\sigma$) C.L. for 2 d.o.f. While the
best-fit point is nominally unchanged by the NSI, the shape of the
$\chi^2$ function is significantly altered, so that small values
of the mixing angle are now much less disfavored.

\section{Conclusion and implications}
In conclusion, the three-flavor oscillation analysis exhibits
qualitatively new features compared with the
$\nu_\mu\leftrightarrow\nu_\tau$ studies. While giving no
positive evidence for non-standard physics, the data do not
exclude large NSI. The old bound $|\eptt|< 0.15$ at 99\% C.L. is
significantly relaxed by the addition of nonzero $\epet$: if
$\epet$ and $\eptt$ are varied simultaneously, even $\eptt\sim 1$
is allowed. The effect is even more dramatic when one compares the
bounds on the flavor-changing interactions: the bound
$\epmt<\mbox{a few}\times 10^{-2}$ \cite{Maltoni2001} does not
extend to $\epet$, for which order one values are allowed.
The bound $\epet < 0.38$ at 99\% C.L. -- found for the case when
the other epsilons vanish -- is likewise new.
We also stress that the effects of these large NSI are absolutely
non-trivial. In particular, the best-fit values of the oscillation
parameters shift to smaller angles and higher $\Delta m^2$.
The physical mechanism behind this is well understood
analytically.

These properties have important implications for other neutrino
experiments. For example, the MINOS experiment will measure the mixing
angle $\theta$ with good precision; assuming the true value
$\theta=\pi/4$ one expects $\sin^2 \theta>0.4$ at 99\% C.L.
\cite{MINOSreach}. Given its baseline of only 735 km, MINOS will
be almost free from matter effects and will therefore determine
the vacuum value of $\theta$. Thus, it can look for the large NSI
effects described here: a measurement of $\sin^2 \theta>0.4$ and
low $\Delta m^2$ would contribute to disfavor large NSI, while in
the opposite case ($\sin^2 \theta\sim 0.3$) one would have
indications of non-standard physics.

Further tests can be performed at future experiments with longer
baseline (a few $\times 10^3$ km) and sufficiently high energy, $E_\nu
\sim 20-30$ GeV. These will be directly sensitive to the NSI matter
effects and a discrepancy between them and MINOS could be a sign of
NSI.

It is worth mentioning that, in principle, another signature of large
NSI could be an unexpected number of neutral current events in a
detector relative to the number of charged current events for a given
oscillation scenario. This effect would be due to the modification of
the \emph{detection} neutral current cross section by large $\epet$
and $\eptt$. Unfortunately, no firm prediction on the size of the
effect can be made, since the effect additionally depends on the
values of the axial couplings that do not enter the propagation
Hamiltonian \footnote{It is for this reason that the NC information
  \cite{SKNC,SNOsalt} was not used in our analysis here and in
  \cite{oursolar}.}.

Finally, there are important implications for solar neutrinos.
Indeed, $\epee$, $\epet$, and $\eptt$ also enter the solar
neutrino survival probability, and may alter the interpretation of
the current data \cite{oursolar,brazilians,valencians}. Our
results here demonstrate that the LMA-0 solution, obtained with
non-zero $\epet$ as in \cite{oursolar}, is compatible with the
atmospheric data.
On the other hand, solar neutrinos, being sensitive to the phase of
$\epet$, may rule out a sizable portion of the parameter space with
$\epet<0$ \cite{oursolar,inprep}. This exclusion is impacted by the latest
results from KamLAND \cite{KamLAND2004} and it would be very
interesting to see the combined bound from the atmospheric, solar, and
KamLAND data.

\begin{acknowledgments}
  We are very grateful to Thomas Schwetz for sharing with us his K2K
  analysis.  We thank the Aspen Center for Physics where part of this
  research was completed for hospitality. A.~F. was supported by the
  Department of Energy, under contract W-7405-ENG-36, C.~L. by a
  grant-in-aid from the W.~M.~Keck Foundation and the NSF grant
  PHY-0070928, and M.~M. by the NSF grant PHY-0354776.
\end{acknowledgments}


\begin{thebibliography}{99}


\bibitem{Maltoni2001}
N.~Fornengo, M.~Maltoni, R.~T.~Bayo and J.~W.~F.~Valle,
Phys.\ Rev.\ D {\bf 65}, 013010 (2002) [arXiv:hep-ph/0108043].

\bibitem{Guzzo2001}
M.~Guzzo, P.~C.~de Holanda, M.~Maltoni, H.~Nunokawa, M.~A.~Tortola
and J.~W.~F.~Valle,
Nucl.\ Phys.\ B {\bf 629}, 479 (2002) [arXiv:hep-ph/0112310].

\bibitem{ConchaMichele2004}
M.~C.~Gonzalez-Garcia and M.~Maltoni,
arXiv:hep-ph/0404085 (to appear in PRD).

\bibitem{oursolar}
A.~Friedland, C.~Lunardini and C.~Pena-Garay,
Phys.\ Lett.\ B {\bf 594}, 347 (2004)
[arXiv:hep-ph/0402266].

\bibitem{brazilians}
M.~M.~Guzzo, P.~C.~de Holanda and O.~L.~G.~Peres,
Phys.\ Lett.\ B {\bf 591}, 1 (2004) [arXiv:hep-ph/0403134].

\bibitem{valencians}
O.~G.~Miranda, M.~A.~Tortola and J.~W.~F.~Valle,
arXiv:hep-ph/0406280.

\bibitem{inprep}
    A.~Friedland, C.~Lunardini, M.~Maltoni, {\it in preparation}.

\bibitem{EngelPLB2000}
R.~Engel, T.~K.~Gaisser and T.~Stanev,
Phys.\ Lett.\ B {\bf 472}, 113 (2000)
[arXiv:hep-ph/9911394].

\bibitem{solareffects}
C.~W.~Kim and U.~W.~Lee,
Phys.\ Lett.\ B {\bf 444}, 204 (1998);
O.~L.~G.~Peres and A.~Y.~Smirnov,
Phys.\ Lett.\ B {\bf 456}, 204 (1999);
M.~C.~Gonzalez-Garcia, M.~Maltoni and A.~Y.~Smirnov,
arXiv:hep-ph/0408170.

\bibitem{nunu}
A.~Friedland and C.~Lunardini,
Phys.\ Rev.\ D {\bf 68}, 013007 (2003);
JHEP {\bf 0310}, 043 (2003).

\bibitem{nutev}
G.~P.~Zeller {\it et al.},
Phys.\ Rev.\ Lett.\  {\bf 88}, 091802 (2002) [Erratum-ibid.\  {\bf
90}, 239902 (2003)] [hep-ex/0110059].

\bibitem{charm}
P.~Vilain {\it et al.},
Phys.\ Lett.\ B {\bf 335}, 246 (1994).

\bibitem{Davidson:2003ha}
S.~Davidson, C.~Pena-Garay, N.~Rius and A.~Santamaria,
JHEP {\bf 0303}, 011 (2003)
[arXiv:hep-ph/0302093].

\bibitem{BerezhianiRossi}
Z.~Berezhiani, A.~Rossi,
Phys.\ Lett.\ B {\bf 535}, 207 (2002).

\bibitem{Hayato:2003} 
  Y. Hayato, Super-Kamiokande Coll., in {\it
    Proceedings of the International Europhysics Conference on High
    Energy Physics}, Aachen, July 2003, [Eur. Phys. J. C 33, s829
  (2004), supplement], 
  \verb"http://eps2003.physik.rwth-aachen.de/"
\verb"transparencies/07/index.php".

\bibitem{Honda:2004yz}
  M.~Honda, T.~Kajita, K.~Kasahara and S.~Midorikawa,
  astro-ph/0404457.


\bibitem{Maltoni:2004ei}
  M.~Maltoni, T.~Schwetz, M.~A.~Tortola and J.~W.~F.~Valle,
  arXiv:hep-ph/0405172.

\bibitem{MINOSreach}
V.~D.~Barger, A.~M.~Gago, D.~Marfatia, W.~J.~C.~Teves, B.~P.~Wood and R.~Zukanovich Funchal,
Phys.\ Rev.\ D {\bf 65}, 053016 (2002)
[arXiv:hep-ph/0110393].

\bibitem{KamLAND2004}
T.~Araki {\it et al.},
arXiv:hep-ex/0406035.

\bibitem{SKNC}
S.~Fukuda {\it et al.},
Phys.\ Rev.\ Lett.\  {\bf 85}, 3999 (2000) [arXiv:hep-ex/0009001];
H. Sobel, in proceedings of Neutrino 2000, XIXth International
Conference on Neutrino Physics and Astrophysics, \verb"http://"
\verb"nu2000.sno.laurentian.ca/H.Sobel/index.html";
 M. Shiozawa,
in proceedings of Neutrino 2002, XXth International Conference on
Neutrino Physics and Astrophysics,
\verb"http://neutrino2002.ph.tum.de/pages/"
\verb"transparencies/shiozawa/index.html".

\bibitem{SNOsalt}
S.~N.~Ahmed {\it et al.}  [SNO Collaboration],
Phys.\ Rev.\ Lett.\  {\bf 92}, 181301 (2004)
[arXiv:nucl-ex/0309004].


\end{thebibliography}
\end{document}